\documentclass[aps,twocolumn, floatfix,superscriptaddress]{revtex4-1}
\usepackage{graphicx}
\usepackage{hyperref}
\usepackage{amsmath}
\usepackage{amsfonts}
\usepackage{mathtools,bm}
\usepackage[amssymb]{SIunits}
\usepackage{color}
\usepackage{mathrsfs}
\usepackage{animate}

\begin{document}
\flushbottom
\title{Finite-energy accelerating beam dynamics in wavelet-based representations}

\author{David Colas}
\email{d.colas@uq.edu.au}
\affiliation{ARC Centre of Excellence in Future Low-Energy Electronics Technologies, School of Mathematics and Physics, University of Queensland, St Lucia, Queensland 4072, Australia}

\author{Fabrice P. Laussy}
\affiliation{Faculty of Science and Engineering, University of Wolverhampton, Wulfruna St, Wolverhampton WV1 1LY, United Kingdom}
\affiliation{Russian Quantum Center, Novaya 100, 143025 Skolkovo, Moscow Region, Russia}

\author{Matthew J. Davis}
\affiliation{ARC Centre of Excellence in Future Low-Energy Electronics Technologies, School of Mathematics and Physics, University of Queensland, St Lucia, Queensland 4072, Australia}

\begin{abstract}
	Accelerating beams are wave packets which appear to spontaneously accelerate without external potentials or applied forces. Since their first physical realisation in the form of Airy beams, they have found applications on various platforms, spanning from optics to plasma physics. We investigate the dynamics of examples of finite-energy accelerating beams derived from catastrophe theory. We use a Madelung transformation in momentum-space, combined with a wavelet transformed analysis, to demonstrate that the beams' properties arise from a special type of vanishing self-interference. We identify the modes responsible for the wave packet's acceleration and we derive the general acceleration for higher-order cupsoid-related beams. We also demonstrate how bright solitons resulting from nonlinear Airy beams can be unambiguously detected using the wavelet transform. This methodology will allow for a better understanding of nontrivial wave packet dynamics.
\end{abstract}

 \date{\today} \maketitle

Accelerating beams are wave packet solutions of the free Schr\"{o}dinger equation which possess the spectacular properties of accelerating without the need of an external potential and without diffracting. They were first discovered in 1979 by Berry and Balazs~\cite{berry79a} with the Airy beam, a non-physical solution that was long considered to be mathematical curiosity.
About three decades later, physical (square-integrable)  approximations of Airy beams were experimentally demonstrated~\cite{siviloglou07a,siviloglou07b,ellenbogen09a}, in which the resulting Airy packets exhibited their special properties for a finite time. Finite-energy Airy beams (FEABs) have since been observed in platforms other than optically-based ones, using \textit{e.g.}, electron beams~\cite{voloch13a} or surface plasmon-polaritons~\cite{zhangP11a}.
These realizations set the ground for a wide range of applications~\cite{baumgartl08a,polynkin09a,vettenburg14a,abdollahpour10a,gu10a,nagar19a}.
  FEABs have also been extensively studied in a nonlinear context. In the presence of a self-focusing nonlinearity, the packet spontaneously splits between a weak accelerating remanent and an ``off-shooting'' bright soliton (BS)~\cite{kaminer11a,lotti11a,zhang13a,zhang14a,zhangL17a,zhangX18a,bouchet18a}, bringing even richer physics and further potential applications.

More fundamentally, Airy beams are only the simplest example of a whole class of caustic beams that arise within the framework of catastrophe theory, introduced Thom~\cite{thom_book72a,berry80a}. Caustic beams emerge from canonical diffraction integrals of codimension $K$
\begin{equation}
\xi_K(\boldsymbol{r})=\int_{-\infty}^{+\infty} \textrm{e}^{i V_K(u;\boldsymbol{r})}du \,,
\end{equation}
with the associated potential functions
\begin{equation}
V_K(u;\boldsymbol{r}) = u^{K+2} + \sum_{n=1}^{K} r_n u^n  \,,
\end{equation}
which depend on one state variable $u$ and on certain control parameters $r_n$. For example, the  fold catastrophe ($K=1$) is related to the Airy function that is the FEAB's building block :
\begin{equation}
\xi_1(x)=\int_{-\infty}^{+\infty} \textrm{e}^{i (u^3 +u x)}du =\frac{2\pi}{\sqrt[3]{3}}\textrm{Ai}\left(\frac{x}{\sqrt[3]{3}}\right) \,,
\end{equation}
while higher-order catastrophes have also been recently studied. The cusp catastrophe ($K=2$) was realised experimentally with Pearcey beams~\cite{ring12a,zangf19a} and the swallowtail catastrophe ($K=3$) with swallowtail beams~\cite{zannotti17a}.

In this letter we study finite-energy accelerating beam dynamics from a new perspective. We employ the wavelet transform (WT), a spectral decomposition which provides broad insights into nontrivial wave packets dynamics~\cite{debnath_book15a}. In the context of Schr\"{o}dinger physics, this technique is particularly suited to detect interference between different wave packets~\cite{baker12a}. It has been recently applied to understand the intriguing phenomenon of wave packet self-interference in exciton-polaritons~\cite{colas16a}, atomic condensates~\cite{colas18a} and to explain the formation of nonlinear X-waves in systems which possess a hyperbolic dispersion~\cite{colas19a}.

In the aforementioned cases, the wave dynamics arise due to the properties of the dispersion relation, such as its curvature or the presence of inflection points~\cite{colas16a,colas18a,colas19a}. Accelerating beams are different. Their short time dynamics result from the intrinsic phase engineering of the initial condition combined with the effect of the parabolic dispersion relation.  At long times the latter becomes dominant, and the packet reshapes into a smooth diffusing one. To understand the resulting complex wave packet phase dynamics, we use a Madelung decomposition of the wave function, not in real-space, but in momentum-space, which is here essential to interpret the results of the WT. We find that the accelerating fringes in the wave packet density result from a transient dynamical self-interference of the packet's internal modes. This allows us to derive the packet acceleration from the WT picture only, without the recourse to heavy calculations or special functions. 
We first study the FEAB, both in the linear and nonlinear regimes, before considering the finite-energy Pearcey beam (FEPB). Finally, we derive a general expression for the acceleration of higher-order cupsoid-related beams. 
\\
\\
We start by introducing our method of analysis for a wave packet evolved with the one-dimensional Schr\"{o}dinger equation, here written in momentum-space:
\begin{equation}
  i \partial_t \psi(k,t)=E(k) \psi(k,t)\,,
\label{eq:Schrod}
\end{equation}
with $\hbar=m=1$ and where the kinetic energy has the usual parabolic dispersion $E(k)=k^2/2$. 
We consider a truncated Airy function as the initial condition, which can  be expressed in momentum-space as
\begin{equation}
\psi_0(k)=\mathscr{F}_{k}[\mathrm{Ai}(b x)\exp((a+i k_0) x)]=\frac{\exp\left(\frac{(a+i (k+k_0))^3}{3 b^3}\right)}{2b \sqrt{2 \pi}}\,.
\label{eq:Airy}
\end{equation}
The parameter $b$ governs the width of the peaks in position-space, while $a$ controls the exponential cut-off of the wave function density to ensure its square-integrability~\cite{note5}. The parameter $k_0$ specifies the momentum of the initial condition. The solution of Eq.~(\ref{eq:Schrod}) is obtained by simple integration: $\psi(k,t)=\psi_0(k)\exp(-ik^2t/2)$.
\begin{figure}[t!]
  \includegraphics[width=\linewidth]{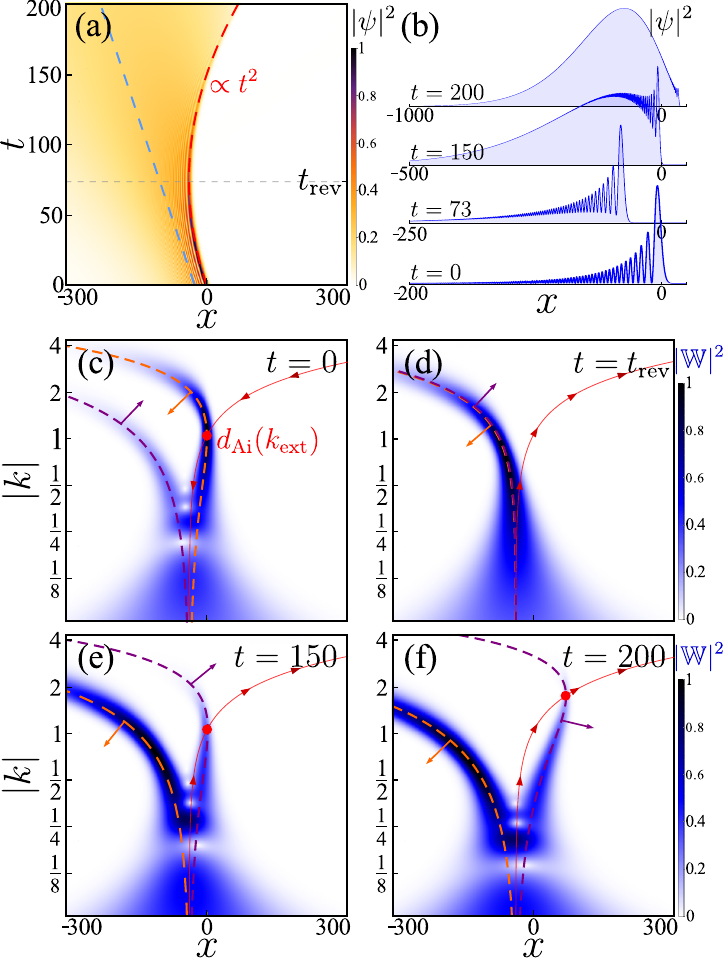}
  \caption{Airy beam propagation. (a) Wave function density $|\psi(x,t)|^2$. The peaks initially move towards the left until $t=t_\mathrm{rev}$. The dashed red, blue and horizontal black lines indicate the wave front trajectory, the packet's center of mass $\bar{x}$, and the reversing time $t_\mathrm{rev}$, respectively. (b) Wave function density $|\psi(x)|^2$ at selected times. (c--f) Corresponding wavelet energy densities $|\mathbb{W}(x,k)|^2$. The dashed purple/orange lines are the mode displacements $d_{\textrm{Ai}}(k,t)$ derived from the FEAB solution's phase (see Eq.~\ref{eq:dk2}). The red dot indicates the position of the branch extremum $d_{\textrm{Ai}}(k_\mathrm{ext})$ around which the self-interference occurs and the solid red line shows its trajectory. Parameters: $a=0.01$, $b=0.3$, $k_0=-1$, $w_\mathcal{G}=8$.}
  \label{fig:1}
\end{figure}
The real-space solution is found by inverse Fourier transform as $\psi(x,t) = \mathscr{F}_{x}^{-1} [\psi(k,t)]$. The space-time dynamics for a FEAB with a negative initial momentum is shown in Fig.~\ref{fig:1}(a), and density profiles at selected times in Fig.~\ref{fig:1}(b). Because of the negative initial ``kick'', the centre of mass of the packet moves to the left (dashed-blue line). However, and as expected, the Airy peaks accelerate along a parabolic trajectory, initially moving to the left, until a reversing time $t_\mathrm{rev}$ after which they continue to accelerate to the right. With zero initial momentum ($k_0=0$), the packet's center of mass would instead remain at the origin ($x=0$) and the peaks would always move to the right.

Other representations of the wave function can also be accessed through the Fourier transform, such as $\psi(k,E)$ (often referred as the \textit{far-field}) or $\psi(x,E)$~\cite{wertz10a}. Alternatively, the WT permits a simultaneous representation of the wave function in both position ($x$) and momentum ($k$). The WT reads~\cite{debnath_book15a}
\begin{equation}
\mathbb{W}(x,k)=(1/\sqrt{|k|})\int_{-\infty}^{+\infty}\psi(x')\mathcal{G}^\ast [(x'-x)/k]\mathrm{d}x'\,.
\end{equation}
For this study we use the Gabor wavelet family
\begin{equation}
\mathcal{G}(x)=\sqrt[4]{\pi}\exp(i w_\mathcal{G} x)\exp(-x^2/2)\,,
\end{equation}
which consists of Gaussian functions with an internal frequency $w_\mathcal{G}$. We apply the WT to the FEAB and show its wavelet energy density $|\mathbb{W}(x,k)|^2$ at four selected times in Fig.~\ref{fig:1}(c--f). It shows two distinct branches that are initially separated at $t=0$.  However, the branches  collapse onto each other at $t_\mathrm{rev}$, before splitting and spreading at longer times. This peculiar distribution can be understood by  analyzing the wave packet's phase dynamics.

It is common to perform a Madelung decomposition of the complex real-space  wave function into an amplitude and a phase term as $\psi(x,t)=\sqrt{N(x,t)}\exp(-i \phi(x,t))$, notably to perform a hydrodynamic analysis~\cite{sonin_book16a}. In this picture, the gradient of the phase corresponds to the fluid velocity $v(x,t) = \partial_x \phi(x,t)$.
Here, we perform the same decomposition to the FEAB, but in momentum-space with $\psi(k,t)=\sqrt{N(k)}\exp(-i \phi(k,t))$, where the amplitude is a time-independent Gaussian of width $\sigma_k=\sqrt{3b^3/4a}$, and where the phase is
\begin{equation}
\phi(k,t)=\frac{(k+k_0)^3 - 3a^2 (k+k_0)}{3 b^3}+\frac{1}{2}k^2 t \,.
\end{equation}

What does the gradient of the phase (with respect to $k$) represent?
For a trivial Gaussian initial condition without any applied phase, the $k$-dependent phase of the packet would be simply $\phi(k,t)=E(k)t$. As the derivative of $E(k)$ gives the group velocity dispersion~\cite{kittel_book96a}, the gradient of the phase now represents a distance as $\partial_k \phi(k,t)=\partial_k E(k) t = v(k) t = k t = d(k,t)$. Explicitly, the term $d(k,t)$ gives the distance traveled by a given mode $k$ after a time $t$. The simple case of the WT for a Gaussian packet is further discussed in the Supplemental Material~\cite{note7}. 

With a FEAB the modes propagate in a more complex fashion, with
\begin{equation}
d_{\textrm{Ai}}(k,t)=\partial_k\phi(k,t)=\frac{(k+k_0)^2 - a^2}{b^3} + k t \,.
\label{eq:dk2}
\end{equation}
From Eq.~(\ref{eq:dk2}) we can see that $d_{\textrm{Ai}}(k,t)$ contains two terms: the first one comes from the FEAB's initial phase, and the second from the dispersion $E(k)$. The FEAB's dynamics arises from the interplay between these two phase terms, which govern the propagation of modes $k$. As only the second term in Eq.~(\ref{eq:dk2}) is time-dependent, at long times, the mode displacement essentially obeys the dispersion relation. 

The mode displacement $d_{\textrm{Ai}}(k,t)$ is plotted in Fig.~\ref{fig:1}(c--f) as dashed-orange/purple lines on top of the wavelet energy density, and show excellent agreement~\cite{note2}. At long times, the mode displacement of the FEAB becomes essentially the one obtained from the dispersion~\cite{note7}.

From the WT analysis, the presence of  fringes in the real-space density can now be understood as self-interference of the wave packet. Indeed, $d_{\textrm{Ai}}(k,t)$ is here a multi-valued function (see Fig.~\ref{fig:1}(c--f)), which leads to a self-interference when the wavelet energy density spreads over its extremum, \textit{i.e.} where it becomes multi-valued. For a given position $x$, two $k$ modes can have support in the wave function and overlap in real-space, resulting in interference. This effect was first identified for condensed-matter systems possessing a non-parabolic dispersion relation, where the extrema correspond to inflection points of the dispersion~\cite{colas16a,colas18a}. 

The trajectory of the branch's extremum point, with coordinates $\{d_{\textrm{Ai}}(k_\mathrm{ext}),k_\mathrm{ext}\}$ in  the $x$-$k$ phase-space, can be determined from the expression of $d_{\textrm{Ai}}(k,t)$. First, solving $\partial_k d_{\textrm{Ai}}(k,t)=0$ for $k$, one obtains $k_\mathrm{ext}=\frac{1}{2}(2 k_0 - b^3 t)$. One can then substitute  $k_\mathrm{ext}$ back into Eq.(\ref{eq:dk2}), which gives:
\begin{equation}
d_{\textrm{Ai}}(k_\mathrm{ext})=\frac{a^2}{b^3} + k_0 t +\frac{b^3 t^2}{4}\,.
\label{eq:parabolicTraj}
\end{equation}
The point $\{d_{\textrm{Ai}}(k_\mathrm{ext}),k_\mathrm{ext}\}$ is shown in Fig.~\ref{fig:1}(c--f) as a red dot and its trajectory as a solid red line. As $d_{\textrm{Ai}}(k_\mathrm{ext})$ corresponds to the largest mode displacement, it gives, from this simple calculation, the trajectory of the FEAB's front wave in real-space. It is indeed parabolic and shown as a dashed-red line in Fig.~\ref{fig:1}(a). One can also obtain the reversal time for the acceleration by solving $\partial_t d_{\textrm{Ai}}(k_\mathrm{ext})=0$ for $t$, which gives $t_\mathrm{rev}=-2 k_0 / b^3$. 

\begin{figure}[t!]
  \includegraphics[width=\linewidth]{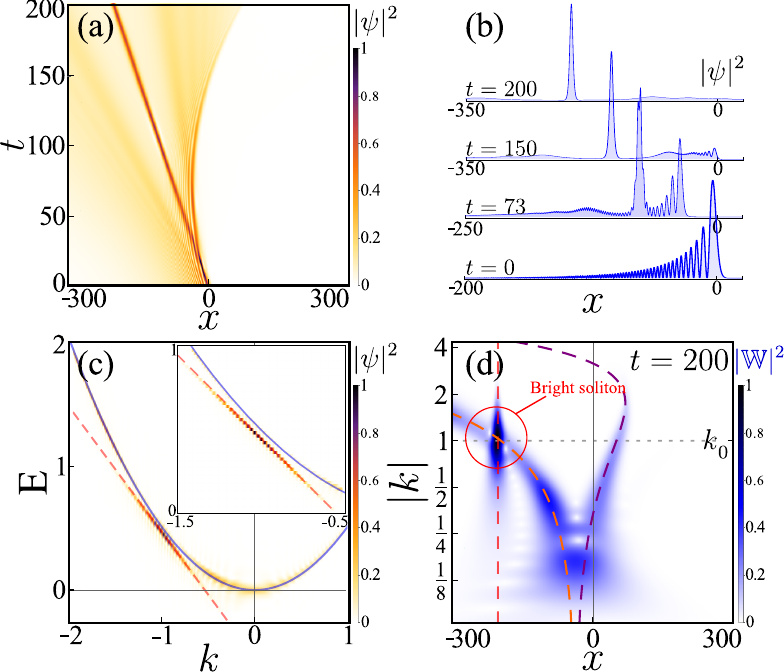}
  \caption{Airy beam propagation with self-focusing nonlinearity $g=-0.5$ and initial momentum $k_0=-1$. (a) Wave function density $|\psi(x,t)|^2$. (b) Wave function density $|\psi(x)|^2$ at selected times. (c) Spectral density $|\psi(k,E)|^2$ lying on the parabolic dispersion $E(k)$ (blue line), and the BS dispersion $E_\mathrm{BS}(k)$ (dashed-red line). Inset: close-up of the BS dispersion. (d) Wavelet energy density $|\mathbb{W}(x,k)|^2$ at $t=200$. The vertical dashed red line stands for the mode displacement associated with the BS dispersion $E_\mathrm{BS}(k)$.}
  \label{fig:2}
\end{figure}

In the systems with a non-parabolic dispersion mentioned earlier~\cite{colas16a,colas18a}, the value $k_\mathrm{ext}$ is time-independent, \textit{i.e.} the self-interference always occurs around the same value of momentum. What makes the FEAB special is the fact that both the coordinates of the point $\{d_{\textrm{Ai}}(k_\mathrm{ext}),k_\mathrm{ext}\}$ around which the self-interference occurs, are time-dependent. This explains why the FEAB's density fringes vanish at long times. One can observe from Fig.~\ref{fig:1}(c--f)  that the wavelet energy density distribution along $k$ is roughly constant in time -- it mostly spreads along $x$. However, the point $\{d_{\textrm{Ai}}(k_\mathrm{ext}),k_\mathrm{ext}\}$ linearly shifts to  large momentum as $k_\mathrm{ext} \propto t$. At long times, there is  then less and less signal available to participate into the self-interference effect, which explains why the wave packet's fringes inevitably disappear. One could maintain the self-interference for a longer time by reducing the exponential cut-off $a$ to increase the spread in momentum-space, since $\sigma_k \propto 1/\sqrt{a}$.

We now briefly consider the effects of introducing an attractive interaction on the  FEAB dynamics, rewriting Eq.~(\ref{eq:Schrod}) as a 1D Gross-Pitaevskii equation:
\begin{equation}
  i \partial_t \psi(x,t)=\left[-\frac{1}{2}\partial_x^2 +g |\psi(x,t)|^2\right]\psi(x,t)\,,
  \label{eq:GPE}
\end{equation}
where $g<0$ is the interaction strength.  We solve Eq.~(\ref{eq:GPE}) for the same parameters as before and show the wave packet dynamics in Fig.~\ref{fig:2}(a,b).

The wave front follows a trajectory that slightly deviates from a parabola~\cite{HuY12a}, along with the generation of a bright peak propagating at constant velocity, previously identified as an ``off-shooting'' BS~\cite{kaminer11a,lotti11a,zhang13a,zhang14a,zhangL17a,zhangX18a,bouchet18a}. This can be confirmed by looking at the \textit{far-field} $|\psi(k,E)|^2$, plotted in Fig.~\ref{fig:2}(c). It follows the usual noninteracting parabolic dispersion, but also displays a clear linear dispersion, tangential to the parabola at the point $k=k_0$, which is an explicit signature of a BS~\cite{egorov09a,sich11a}. The BS dispersion is obtained by performing a Taylor expansion of the main branch around $k_0$, giving $E_\mathrm{BS}(k)=k k_0-k_0^2/2$ .
The BS velocity can be determined from the slope of the linear dispersion as $v_\mathrm{BS}=\partial_k E_\mathrm{BS}(k) =k_0$. In the wavelet analysis, this corresponds to a single mode displacement:
\begin{equation}
d_\mathrm{BS}(k,t)=\partial_k E_\mathrm{BS}(k)t= k_0 t\,,
\label{eq:8}
\end{equation}
which appears as a vertical line in the $x$-$k$ representation, as shown in Fig.~\ref{fig:2}(d). As the BS arises from a non-diffusing mode, \textit{i.e.} a linear dispersion, its wavelet energy density remains localised around the point $\{k_0 t, k_0\}$. This illustrates how the WT can be used to detect the BS signature, complementary to the usual Fourier techniques.
\begin{figure}[t!]
  \includegraphics[width=\linewidth]{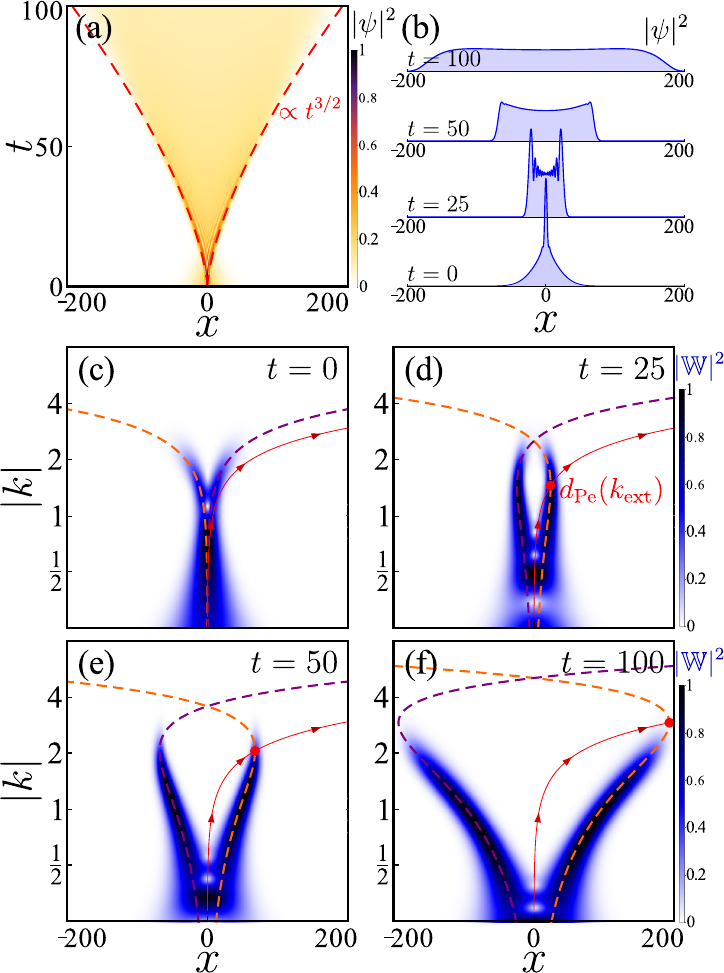}
  \caption{Pearcey beam propagation. (a) Wave function density $|\psi(x,t)|^2$. (b) Wave function density $|\psi(x)|^2$ at selected times. (c--f) Corresponding wavelet energy densities $|\mathbb{W}(x,k)|^2$. The dashed purple/orange lines are the mode displacements $d_{\textrm{Pe}}(k,t)$ derived from Eq.~\ref{eq:dk2}). The red dot indicates the position of one the branch extremum $d_{\textrm{Pe}}(k_\mathrm{ext})$ around which the self-interference occurs and the solid red line shows its trajectory. Parameters: $\sigma=1/1000$.}
  \label{fig:3}
\end{figure}

We now consider the case of the FEPB obtained from the cusp catastrophe ($K=2$):
\begin{equation}
\xi_2(x,y)=\int_{-\infty}^{+\infty} \textrm{e}^{i (u^4 +u^2 y +u x)}du =\textrm{Pe}(x,y)\,,
\end{equation}
that defines the Pearcey function~\cite{note8}. The initial condition for Eq.~\ref{eq:Schrod} is now, in 1D, $\psi_0(x)=\textrm{Pe}(x,0)\exp(-\sigma x^2)$. The square-integrability is ensured by the Gaussian term of width $\sigma$ which cuts off the ``fat'' tails of $\textrm{Pe}(x,0)$.
The Pearcey function can be expressed in momentum-space as $\mathscr{F}_{k} [\textrm{Pe}(x,0)]=2\pi \exp(i k^4)$~\cite{note9}. The full solution of Eq.~\ref{eq:Schrod} for a FEPB reads $\psi(k,t)=2 \pi \exp\left[-i (-k^4 +k^2t/2)\right]$, and typical Pearcey beam dynamics is shown in Fig.~\ref{fig:3}(a,b). Following the procedure previously applied to the Airy beam, one obtains the mode propagation distance $d_{\textrm{Pe}}(k,t)= -4 k^3 + kt$ and hence the wave front trajectory $k_\mathrm{ext} = \pm \sqrt{t/12}$. Subsequently, the propagation distance for the largest mode is
\begin{equation}
d_{\textrm{Pe}}(k_\mathrm{ext})= \pm (t/3)^{3/2} \,.
\label{eq:dkext_Pearcey}
\end{equation}
A double self-interference effect here takes place as $d_{\textrm{Pe}}(k,t)$ is multi-valued for both branches, the interference disappears as the points $\{d_{\textrm{Pe}}(k_\mathrm{ext}),k_\mathrm{ext}\}$ drift away from the wavelet energy density, as shown in Fig.~\ref{fig:3}(c--f).

Unlike the FEAB (Eq.~\ref{eq:parabolicTraj}), the two FEPB's wave fronts accelerate as $t^{3/2}$ and not as $t^2$, which is a consequence of the phase factor imprinted in the initial condition: $\sim\exp(i k^4)$ for the FEPB vs $\sim\exp(i k^3)$ for the FEAB. We can generalise our previous result for an initial phase containing any power of $k$. Considering a packet with a phase proportional to $\exp(i k^n)$, one can find the generalised wave front acceleration 
\begin{equation}
d(k_\mathrm{ext};n) = n(n-2)\left( \frac{t}{n(n-1)}\right)^\frac{1-n}{2-n}\,.
\end{equation}
which is a $t^2$ and $t^{3/2}$ acceleration for $n=3$ and $4$. In the limit of large $n$, one reaches a non-accelerating limit as $d(k_\mathrm{ext};n\rightarrow \infty)=t$.
\\
\\
In conclusion, we have shown that the dynamics of finite-energy accelerating beams can be fully understood from a careful phase dynamics analysis using the WT and a Madelung decomposition. We have identified that the key properties arise from a transient self-interference of the wave packet and that the wave front acceleration can be obtained analytically. We find that the reshaping mechanism originates from a dynamical shift of the extremum mode, that controls the accelerating wave front, towards of regions of low spectral density. This method of analysis is applicable to other accelerating beams with a different phase engineering, as well as nonlinear objects like BS with a well-defined dispersion. 
Supplemental movies S1 to S3 provide an animation of $|\psi(x,t)|^2$ and $|\mathbb{W}(x,k)|^2$ for Figs.~\ref{fig:1} to \ref{fig:3}~\cite{note7}.
\\
\\
\textit{Acknowledgments--}
This research was supported by the Australian Research Council Centre of Excellence in Future Low-Energy Electronics Technologies (project number CE170100039). It was also supported by the Ministry of Science and Education of the Russian Federation through the Russian-Greek project RFMEFI61617X0085 and the Spanish MINECO under contract FIS2015-64951-R (CLAQUE).

\pagebreak

\onecolumngrid
\begin{center}
  \textbf{\large Supplemental Material:\\
Finite-energy accelerating beam dynamics in wavelet-based representations}\\[.2cm]
  David Colas,$^{1,*}$ Fabrice P. Laussy,$^{2,3}$ and Matthew J. Davis$^1$\\[.1cm]
  {\itshape ${}^1$ARC Centre of Excellence in Future Low-Energy Electronics Technologies,\\ School of Mathematics and Physics, University of Queensland, St Lucia, Queensland 4072, Australia\\
  ${}^2$Faculty of Science and Engineering, University of Wolverhampton,\\Wulfruna St, Wolverhampton WV1 1LY, United Kingdom\\
  ${}^3$Russian Quantum Center, Novaya 100, 143025 Skolkovo, Moscow Region, Russia\\}
  ${}^*$Electronic address: d.colas@uq.edu.au\\
(Dated: \today)\\[1cm]
\end{center}
\twocolumngrid

\setcounter{equation}{0}
\setcounter{figure}{0}
\setcounter{table}{0}
\setcounter{page}{1}
\renewcommand{\theequation}{S\arabic{equation}}
\renewcommand{\thefigure}{S\arabic{figure}}
\renewcommand{\bibnumfmt}[1]{[S#1]}
\renewcommand{\citenumfont}[1]{S#1}

In this Supplemental Material we apply the wavelet analysis to the simple case of a diffusing Gaussian wave packet with a parabolic dispersion relation. We discuss the case of an Airy beam where additional damping and amplification are used to dynamically engineer the truncated Airy beam. Finally, we describe the content of the Supplementary Videos. Equations and Figures from the main text are here quoted with numbers, whereas those from the Supplementary Material are prefixed by ``S''.
\subsection{Gaussian wave packet}
 We begin with the one-dimensional Schr\"{o}dinger equation, written in momentum-space
\begin{equation}
  i \partial_t \psi(k,t)=E(k) \psi(k,t)\,,
\label{eq:Schrod}
\end{equation}
where the kinetic energy has the parabolic dispersion $E(k)=k^2/2$.
Taking a Gaussian wave packet as the initial condition for Eq.~(\ref{eq:Schrod}), which can be either written in real or momentum space as $\psi_0(k)=\mathscr{F}_{k}[\exp(-x^2/2\sigma_x^2)]\simeq \exp(-k^2/2\sigma_k^2)$, with $\sigma_x=1/\sigma_k$, the full solution in momentum-space is
\begin{equation}
\psi(k,t)=\exp\left(-\frac{i k^2 t}{2}\right)\exp\left(-\frac{k^2}{2\sigma_k^2}\right)\,.
\label{eq:solSchrod2}
\end{equation}

\begin{figure}[t!]
  \includegraphics[width=\linewidth]{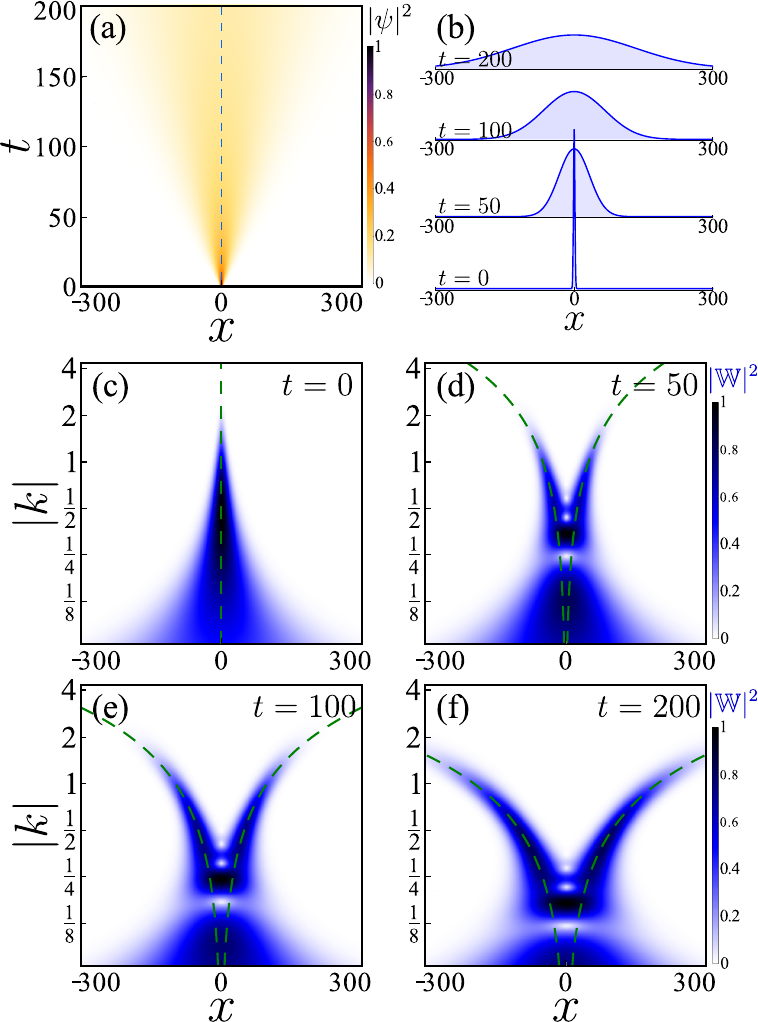}
  \caption{Diffusion of an initially sharp ($\sigma=1$) Gaussian wave packet. (a) Wave function density $|\psi(x,t)|^2$. The dashed blue line indicates the packet's center of mass $\bar{x}$. (b) Density profile at selected times. (c--f) Corresponding wavelet energy density $|\mathbb{W}(x,k)|^2$. The dashed green lines indicate the  displacement $d(k)$ of mode $k$ derived from the parabolic dispersion relation. Supplemental Video S4 provides an animation of the Gaussian packet dynamics with its WT.}
  \label{fig:0}
\end{figure}

In Fig.~\ref{fig:0}(a) we show an example of the well-known freely diffusing Gaussian wave packet, initialized with $\sigma_x = 1$, as seen in the density profiles in Fig.~\ref{fig:0}(b). The dashed-blue line here shows the position of the packet's center of mass.

We apply the WT to the diffusing Gaussian packet and show the wavelet energy density $|\mathbb{W}(x,k)|^2$ at selected times in Fig.~\ref{fig:0}(c--f). It is initially tightly distributed around $x=0$ where the packet stands, and then spreads as two branches.

As in the main text, we decompose the complex momentum-space wave function into an amplitude term and a phase term as $\psi(k,t)=\sqrt{N(k)}\exp(-i \phi(k,t))$, with the amplitude being
\begin{equation}
\sqrt{N(k)}=\exp\left(-\frac{k^2}{2\sigma_k^2}\right)\,,
\end{equation}
and the phase
\begin{equation}
\phi(k,t)=E(k)t = k^2 t/2\,.
\end{equation}
We now compute the gradient of the phase (with respect to $k$):
\begin{equation}
d(k,t)=\partial_k \phi(k,t)=\partial_k E(k) t = v(k) t = k t\,.
\label{eq:dk1}
\end{equation}
As the $k$-dependent velocity is obtained by taking the derivative of the dispersion relation, the gradient of the phase in momentum-space represents a distance $d(k,t)$ of propagation of a given mode $k$ at a time $t$. The distance travelled for each mode of the wave packet is superimposed on the wavelet energy density shown in Fig.~\ref{fig:0}(c--f). We note that this result would be qualitatively comparable to other non-Gaussian wave packets evolved on the same parabolic dispersion, as long as they do not initially contain any complex phase relationships. 
This case corresponds to the large times limit of finite-energy Airy beams shown in Fig.~[1].
\subsection{Dynamical mode filtering/amplification}
\begin{figure}[t!]
  \includegraphics[width=\linewidth]{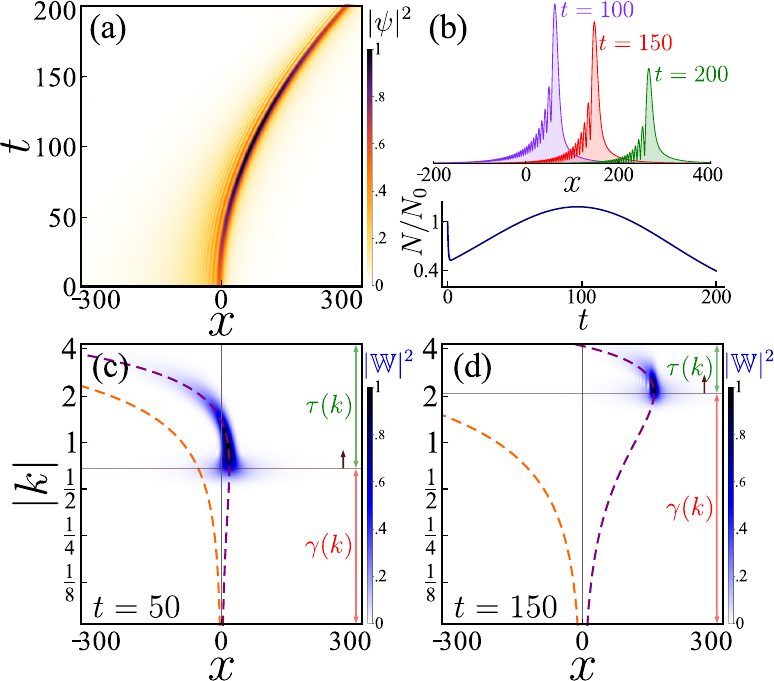}
  \caption{ Airy-engineered wave front. (a) Wave function density $|\psi(x,t)|^2$. (b) Wave function density $|\psi(x)|^2$ at selected times (top) and evolution of the total normalised population (bottom). (c,d) Wavelet energy density $|\mathbb{W}(x,k)|^2$ at two selected times. The left branch is fully damped from $t=0$ and the right branch is progressively damped, following the point $\{d(k_\mathrm{ext}),k_\mathrm{ext}\}$, while the rest of the signal is amplified. Supplemental Video S5 provides an animation of the engineered-FEAB dynamics with its WT.}
  \label{fig:2}
\end{figure}
From the wavelet spectra in Fig.~S1 it can be seen that, at long times, the energy density around $\{d_\mathrm{Ai}(k_\mathrm{ext}),k_\mathrm{ext}\}$ is small. The majority of the signal comes from modes participating in the reshaping of the packet into a Gaussian. Therefore we consider damping out those modes, and enhancing those contributing to the self-interference, and hence to the accelerating peaks. 
We set up a dynamical high-pass filter to damp all the modes below a certain momentum, close to $k_\mathrm{ext}$. This can be translated as a momentum and time-dependent loss term for the Schr\"{o}dinger equation. Similarly, we set a dynamical amplification for the remaining modes with a gain term, in order to limit the decay of the normalistion. We can now rewrite Eq.~(\ref{eq:Schrod}) as an open-dissipative Schr\"{o}dinger equation
\begin{equation}
  i \partial_t \psi(k,t)=\left[\frac{k^2}{2} +i\left(\tau(k,t) -\gamma(k,t)\right) \right] \psi(k,t)\,,
\label{eq:SchrodOD}
\end{equation}
where $\tau$ and $\gamma$ are the gain and loss terms, respectively. We define these functions as
\begin{align} 
\gamma(k,t)=\Theta[-k+\frac{1}{2}b^3t]\,, \\ 
\tau(k,t)= \beta \Theta[k-\frac{1}{2}b^3t] \,,
\end{align}
where $\Theta$ is a Heavyside step function and $\beta$ a constant chosen to control the amplification of the remaining modes. The ``boundary'' in $k$-space between damping and amplification here follows the position of $d_\mathrm{Ai}(k_\mathrm{ext})$ which is a linear function of time. 

To solve Eq.~(\ref{eq:SchrodOD}), we use the same parameters as in Fig.~S1 (but with $k_0=0$). The newly Airy-engineered wave packet is shown in Fig.~\ref{fig:2}(a,b) and does not display any Gaussian reshaping. Instead, it is an essentially shape-preserving and accelerating wave front. The normalisation is approximately constant over the considered time interval due to the amplification of the high $k$ modes. The effect of the high-pass filter is shown in Fig.~\ref{fig:2}(c,d). The signal overlapping the left branch of $d(k,t)$ is fully damped from $t=0$, leading to a sudden population decrease at short times. The signal overlapping the right branch is then progressively damped, following the drift of $d_\mathrm{Ai}(k_\mathrm{ext})$, while the remaining signal is amplified.

This is the simplest way to dynamically damp and amplify a desired range of modes. This is sufficient to prevent the total population varying by  more than a factor of two over the time interval we consider. This procedure could be further optimised using more complex functions for $\tau(k,t)$ and $\gamma(k,t)$, notably to enable an experimental realization of this scheme.

\subsection{Supplementary videos}
Five videos are provided showing animations of Figs.~1--3 of the main text, and Fig.~\ref{fig:0} and \ref{fig:2} of the Supplementary Material, respectively.  In  Supplementary Videos S1 and S3, corresponding to Fig.~1 and 3, we report the dynamical position of the point $\{d(k_\mathrm{ext}),k_\mathrm{ext}\}$. The displacement $d(k_\mathrm{ext})$ corresponds to the position of the main peak in the real-space density, and it is indicated a solid vertical red line.

\begin{thebibliography}{40}%
\makeatletter
\providecommand \@ifxundefined [1]{%
 \@ifx{#1\undefined}
}%
\providecommand \@ifnum [1]{%
 \ifnum #1\expandafter \@firstoftwo
 \else \expandafter \@secondoftwo
 \fi
}%
\providecommand \@ifx [1]{%
 \ifx #1\expandafter \@firstoftwo
 \else \expandafter \@secondoftwo
 \fi
}%
\providecommand \natexlab [1]{#1}%
\providecommand \enquote  [1]{``#1''}%
\providecommand \bibnamefont  [1]{#1}%
\providecommand \bibfnamefont [1]{#1}%
\providecommand \citenamefont [1]{#1}%
\providecommand \href@noop [0]{\@secondoftwo}%
\providecommand \href [0]{\begingroup \@sanitize@url \@href}%
\providecommand \@href[1]{\@@startlink{#1}\@@href}%
\providecommand \@@href[1]{\endgroup#1\@@endlink}%
\providecommand \@sanitize@url [0]{\catcode `\\12\catcode `\$12\catcode
  `\&12\catcode `\#12\catcode `\^12\catcode `\_12\catcode `\%12\relax}%
\providecommand \@@startlink[1]{}%
\providecommand \@@endlink[0]{}%
\providecommand \url  [0]{\begingroup\@sanitize@url \@url }%
\providecommand \@url [1]{\endgroup\@href {#1}{\urlprefix }}%
\providecommand \urlprefix  [0]{URL }%
\providecommand \Eprint [0]{\href }%
\providecommand \doibase [0]{http://dx.doi.org/}%
\providecommand \selectlanguage [0]{\@gobble}%
\providecommand \bibinfo  [0]{\@secondoftwo}%
\providecommand \bibfield  [0]{\@secondoftwo}%
\providecommand \translation [1]{[#1]}%
\providecommand \BibitemOpen [0]{}%
\providecommand \bibitemStop [0]{}%
\providecommand \bibitemNoStop [0]{.\EOS\space}%
\providecommand \EOS [0]{\spacefactor3000\relax}%
\providecommand \BibitemShut  [1]{\csname bibitem#1\endcsname}%
\let\auto@bib@innerbib\@empty
\bibitem [{\citenamefont {Berry}\ and\ \citenamefont
  {Balazs}(1979)}]{berry79a}%
  \BibitemOpen
  \bibfield  {author} {\bibinfo {author} {\bibfnamefont {M.~V.}\ \bibnamefont
  {Berry}}\ and\ \bibinfo {author} {\bibfnamefont {N.~L.}\ \bibnamefont
  {Balazs}},\ }\href@noop {} {\bibfield  {journal} {\bibinfo  {journal} {Am. J.
  Phys.}\ }\textbf {\bibinfo {volume} {47}},\ \bibinfo {pages} {264} (\bibinfo
  {year} {1979})}\BibitemShut {NoStop}%
\bibitem [{\citenamefont {Siviloglou}\ \emph {et~al.}(2007)\citenamefont
  {Siviloglou}, \citenamefont {Broky}, \citenamefont {Dogariu},\ and\
  \citenamefont {Christodoulides}}]{siviloglou07a}%
  \BibitemOpen
  \bibfield  {author} {\bibinfo {author} {\bibfnamefont {G.~A.}\ \bibnamefont
  {Siviloglou}}, \bibinfo {author} {\bibfnamefont {J.}~\bibnamefont {Broky}},
  \bibinfo {author} {\bibfnamefont {A.}~\bibnamefont {Dogariu}}, \ and\
  \bibinfo {author} {\bibfnamefont {D.~N.}\ \bibnamefont {Christodoulides}},\
  }\href@noop {} {\bibfield  {journal} {\bibinfo  {journal} {Phys. Rev. Lett.}\
  }\textbf {\bibinfo {volume} {99}},\ \bibinfo {pages} {213901} (\bibinfo
  {year} {2007})}\BibitemShut {NoStop}%
\bibitem [{\citenamefont {Siviloglou}\ and\ \citenamefont
  {Christodoulides}(2007)}]{siviloglou07b}%
  \BibitemOpen
  \bibfield  {author} {\bibinfo {author} {\bibfnamefont {G.~A.}\ \bibnamefont
  {Siviloglou}}\ and\ \bibinfo {author} {\bibfnamefont {D.~N.}\ \bibnamefont
  {Christodoulides}},\ }\href@noop {} {\bibfield  {journal} {\bibinfo
  {journal} {Opt. Lett.}\ }\textbf {\bibinfo {volume} {32}},\ \bibinfo {pages}
  {979} (\bibinfo {year} {2007})}\BibitemShut {NoStop}%
\bibitem [{\citenamefont {Ellenbogen}\ \emph {et~al.}(2009)\citenamefont
  {Ellenbogen}, \citenamefont {Voloch-Bloch}, \citenamefont {Ganany-Padowicz},\
  and\ \citenamefont {Arie}}]{ellenbogen09a}%
  \BibitemOpen
  \bibfield  {author} {\bibinfo {author} {\bibfnamefont {T.}~\bibnamefont
  {Ellenbogen}}, \bibinfo {author} {\bibfnamefont {N.}~\bibnamefont
  {Voloch-Bloch}}, \bibinfo {author} {\bibfnamefont {A.}~\bibnamefont
  {Ganany-Padowicz}}, \ and\ \bibinfo {author} {\bibfnamefont {A.}~\bibnamefont
  {Arie}},\ }\href@noop {} {\bibfield  {journal} {\bibinfo  {journal} {Nat.
  Photon.}\ }\textbf {\bibinfo {volume} {3}},\ \bibinfo {pages} {395} (\bibinfo
  {year} {2009})}\BibitemShut {NoStop}%
\bibitem [{\citenamefont {Voloch-Bloch}\ \emph {et~al.}(2013)\citenamefont
  {Voloch-Bloch}, \citenamefont {Lereah}, \citenamefont {Lilach}, \citenamefont
  {Gover},\ and\ \citenamefont {Arie}}]{voloch13a}%
  \BibitemOpen
  \bibfield  {author} {\bibinfo {author} {\bibfnamefont {N.}~\bibnamefont
  {Voloch-Bloch}}, \bibinfo {author} {\bibfnamefont {Y.}~\bibnamefont
  {Lereah}}, \bibinfo {author} {\bibfnamefont {Y.}~\bibnamefont {Lilach}},
  \bibinfo {author} {\bibfnamefont {A.}~\bibnamefont {Gover}}, \ and\ \bibinfo
  {author} {\bibfnamefont {A.}~\bibnamefont {Arie}},\ }\href@noop {} {\bibfield
   {journal} {\bibinfo  {journal} {Nature}\ }\textbf {\bibinfo {volume}
  {494}},\ \bibinfo {pages} {331} (\bibinfo {year} {2013})}\BibitemShut
  {NoStop}%
\bibitem [{\citenamefont {Zhang}\ \emph {et~al.}(2011)\citenamefont {Zhang},
  \citenamefont {Wang}, \citenamefont {Liu}, \citenamefont {Yin}, \citenamefont
  {Lu}, \citenamefont {Chen},\ and\ \citenamefont {Zhang}}]{zhangP11a}%
  \BibitemOpen
  \bibfield  {author} {\bibinfo {author} {\bibfnamefont {P.}~\bibnamefont
  {Zhang}}, \bibinfo {author} {\bibfnamefont {S.}~\bibnamefont {Wang}},
  \bibinfo {author} {\bibfnamefont {Y.}~\bibnamefont {Liu}}, \bibinfo {author}
  {\bibfnamefont {X.}~\bibnamefont {Yin}}, \bibinfo {author} {\bibfnamefont
  {C.}~\bibnamefont {Lu}}, \bibinfo {author} {\bibfnamefont {Z.}~\bibnamefont
  {Chen}}, \ and\ \bibinfo {author} {\bibfnamefont {X.}~\bibnamefont {Zhang}},\
  }\href@noop {} {\bibfield  {journal} {\bibinfo  {journal} {Opt. Lett.}\
  }\textbf {\bibinfo {volume} {36}},\ \bibinfo {pages} {3191} (\bibinfo {year}
  {2011})}\BibitemShut {NoStop}%
\bibitem [{\citenamefont {Baumgartl}\ \emph {et~al.}(2008)\citenamefont
  {Baumgartl}, \citenamefont {Mazilu},\ and\ \citenamefont
  {Dholakia}}]{baumgartl08a}%
  \BibitemOpen
  \bibfield  {author} {\bibinfo {author} {\bibfnamefont {J.}~\bibnamefont
  {Baumgartl}}, \bibinfo {author} {\bibfnamefont {M.}~\bibnamefont {Mazilu}}, \
  and\ \bibinfo {author} {\bibfnamefont {K.}~\bibnamefont {Dholakia}},\
  }\href@noop {} {\bibfield  {journal} {\bibinfo  {journal} {Nat. Photon.}\
  }\textbf {\bibinfo {volume} {2}},\ \bibinfo {pages} {675} (\bibinfo {year}
  {2008})}\BibitemShut {NoStop}%
\bibitem [{\citenamefont {Polynkin}\ \emph {et~al.}(2009)\citenamefont
  {Polynkin}, \citenamefont {Kolesik}, \citenamefont {Moloney}, \citenamefont
  {Siviloglou},\ and\ \citenamefont {Christodoulides}}]{polynkin09a}%
  \BibitemOpen
  \bibfield  {author} {\bibinfo {author} {\bibfnamefont {P.}~\bibnamefont
  {Polynkin}}, \bibinfo {author} {\bibfnamefont {M.}~\bibnamefont {Kolesik}},
  \bibinfo {author} {\bibfnamefont {J.~V.}\ \bibnamefont {Moloney}}, \bibinfo
  {author} {\bibfnamefont {G.~A.}\ \bibnamefont {Siviloglou}}, \ and\ \bibinfo
  {author} {\bibfnamefont {D.~N.}\ \bibnamefont {Christodoulides}},\
  }\href@noop {} {\bibfield  {journal} {\bibinfo  {journal} {Science}\ }\textbf
  {\bibinfo {volume} {324}},\ \bibinfo {pages} {229} (\bibinfo {year}
  {2009})}\BibitemShut {NoStop}%
\bibitem [{\citenamefont {Vettenburg}\ \emph {et~al.}(2014)\citenamefont
  {Vettenburg}, \citenamefont {Dalgarno}, \citenamefont {Nylk}, \citenamefont
  {Coll-Llado}, \citenamefont {Ferrier}, \citenamefont {Cizmar}, \citenamefont
  {Gunn-Moore},\ and\ \citenamefont {Dholakia}}]{vettenburg14a}%
  \BibitemOpen
  \bibfield  {author} {\bibinfo {author} {\bibfnamefont {T.}~\bibnamefont
  {Vettenburg}}, \bibinfo {author} {\bibfnamefont {H.~I.~C.}\ \bibnamefont
  {Dalgarno}}, \bibinfo {author} {\bibfnamefont {J.}~\bibnamefont {Nylk}},
  \bibinfo {author} {\bibfnamefont {C.}~\bibnamefont {Coll-Llado}}, \bibinfo
  {author} {\bibfnamefont {D.~E.~K.}\ \bibnamefont {Ferrier}}, \bibinfo
  {author} {\bibfnamefont {T.}~\bibnamefont {Cizmar}}, \bibinfo {author}
  {\bibfnamefont {F.~J.}\ \bibnamefont {Gunn-Moore}}, \ and\ \bibinfo {author}
  {\bibfnamefont {K.}~\bibnamefont {Dholakia}},\ }\href@noop {} {\bibfield
  {journal} {\bibinfo  {journal} {Nat. Methods}\ }\textbf {\bibinfo {volume}
  {11}},\ \bibinfo {pages} {541} (\bibinfo {year} {2014})}\BibitemShut
  {NoStop}%
\bibitem [{\citenamefont {Abdollahpour}\ \emph {et~al.}(2010)\citenamefont
  {Abdollahpour}, \citenamefont {Suntsov}, \citenamefont {Papazoglou},\ and\
  \citenamefont {Tzortzakis}}]{abdollahpour10a}%
  \BibitemOpen
  \bibfield  {author} {\bibinfo {author} {\bibfnamefont {D.}~\bibnamefont
  {Abdollahpour}}, \bibinfo {author} {\bibfnamefont {S.}~\bibnamefont
  {Suntsov}}, \bibinfo {author} {\bibfnamefont {D.~G.}\ \bibnamefont
  {Papazoglou}}, \ and\ \bibinfo {author} {\bibfnamefont {S.}~\bibnamefont
  {Tzortzakis}},\ }\href@noop {} {\bibfield  {journal} {\bibinfo  {journal}
  {Phys. Rev. Lett.}\ }\textbf {\bibinfo {volume} {105}},\ \bibinfo {pages}
  {253901} (\bibinfo {year} {2010})}\BibitemShut {NoStop}%
\bibitem [{\citenamefont {Gu}\ and\ \citenamefont {Gbur}(2010)}]{gu10a}%
  \BibitemOpen
  \bibfield  {author} {\bibinfo {author} {\bibfnamefont {Y.}~\bibnamefont
  {Gu}}\ and\ \bibinfo {author} {\bibfnamefont {G.}~\bibnamefont {Gbur}},\
  }\href@noop {} {\bibfield  {journal} {\bibinfo  {journal} {Opt. Lett.}\
  }\textbf {\bibinfo {volume} {35}},\ \bibinfo {pages} {3456} (\bibinfo {year}
  {2010})}\BibitemShut {NoStop}%
\bibitem [{\citenamefont {Nagar}\ and\ \citenamefont
  {Roichman}(2019)}]{nagar19a}%
  \BibitemOpen
  \bibfield  {author} {\bibinfo {author} {\bibfnamefont {H.}~\bibnamefont
  {Nagar}}\ and\ \bibinfo {author} {\bibfnamefont {Y.}~\bibnamefont
  {Roichman}},\ }\href@noop {} {\bibfield  {journal} {\bibinfo  {journal} {Opt.
  Lett.}\ }\textbf {\bibinfo {volume} {44}},\ \bibinfo {pages} {1896} (\bibinfo
  {year} {2019})}\BibitemShut {NoStop}%
\bibitem [{\citenamefont {Kaminer}\ \emph {et~al.}(2011)\citenamefont
  {Kaminer}, \citenamefont {Segev},\ and\ \citenamefont
  {Christodoulides}}]{kaminer11a}%
  \BibitemOpen
  \bibfield  {author} {\bibinfo {author} {\bibfnamefont {I.}~\bibnamefont
  {Kaminer}}, \bibinfo {author} {\bibfnamefont {M.}~\bibnamefont {Segev}}, \
  and\ \bibinfo {author} {\bibfnamefont {D.~N.}\ \bibnamefont
  {Christodoulides}},\ }\href@noop {} {\bibfield  {journal} {\bibinfo
  {journal} {Phys. Rev. Lett.}\ }\textbf {\bibinfo {volume} {106}},\ \bibinfo
  {pages} {213903} (\bibinfo {year} {2011})}\BibitemShut {NoStop}%
\bibitem [{\citenamefont {Lotti}\ \emph {et~al.}(2011)\citenamefont {Lotti},
  \citenamefont {Faccio}, \citenamefont {Couairon}, \citenamefont {Papazoglou},
  \citenamefont {Panagiotopoulos}, \citenamefont {Abdollahpour},\ and\
  \citenamefont {Tzortzakis}}]{lotti11a}%
  \BibitemOpen
  \bibfield  {author} {\bibinfo {author} {\bibfnamefont {A.}~\bibnamefont
  {Lotti}}, \bibinfo {author} {\bibfnamefont {D.}~\bibnamefont {Faccio}},
  \bibinfo {author} {\bibfnamefont {A.}~\bibnamefont {Couairon}}, \bibinfo
  {author} {\bibfnamefont {D.~G.}\ \bibnamefont {Papazoglou}}, \bibinfo
  {author} {\bibfnamefont {P.}~\bibnamefont {Panagiotopoulos}}, \bibinfo
  {author} {\bibfnamefont {D.}~\bibnamefont {Abdollahpour}}, \ and\ \bibinfo
  {author} {\bibfnamefont {S.}~\bibnamefont {Tzortzakis}},\ }\href@noop {}
  {\bibfield  {journal} {\bibinfo  {journal} {Phys. Rev. A}\ }\textbf {\bibinfo
  {volume} {84}},\ \bibinfo {pages} {021807} (\bibinfo {year}
  {2011})}\BibitemShut {NoStop}%
\bibitem [{\citenamefont {Zhang}\ \emph {et~al.}(2013)\citenamefont {Zhang},
  \citenamefont {Beli\'c}, \citenamefont {Wu}, \citenamefont {Zheng},
  \citenamefont {Lu}, \citenamefont {Li},\ and\ \citenamefont
  {Zhang}}]{zhang13a}%
  \BibitemOpen
  \bibfield  {author} {\bibinfo {author} {\bibfnamefont {Y.}~\bibnamefont
  {Zhang}}, \bibinfo {author} {\bibfnamefont {M.}~\bibnamefont {Beli\'c}},
  \bibinfo {author} {\bibfnamefont {Z.}~\bibnamefont {Wu}}, \bibinfo {author}
  {\bibfnamefont {H.}~\bibnamefont {Zheng}}, \bibinfo {author} {\bibfnamefont
  {K.}~\bibnamefont {Lu}}, \bibinfo {author} {\bibfnamefont {Y.}~\bibnamefont
  {Li}}, \ and\ \bibinfo {author} {\bibfnamefont {Y.}~\bibnamefont {Zhang}},\
  }\href@noop {} {\bibfield  {journal} {\bibinfo  {journal} {Opt. Lett.}\
  }\textbf {\bibinfo {volume} {22}},\ \bibinfo {pages} {4585} (\bibinfo {year}
  {2013})}\BibitemShut {NoStop}%
\bibitem [{\citenamefont {Zhang}\ \emph {et~al.}(2014)\citenamefont {Zhang},
  \citenamefont {Beli\'c}, \citenamefont {Zheng}, \citenamefont {Chen},
  \citenamefont {Li}, \citenamefont {Li},\ and\ \citenamefont
  {Zhang}}]{zhang14a}%
  \BibitemOpen
  \bibfield  {author} {\bibinfo {author} {\bibfnamefont {Y.}~\bibnamefont
  {Zhang}}, \bibinfo {author} {\bibfnamefont {M.~R.}\ \bibnamefont {Beli\'c}},
  \bibinfo {author} {\bibfnamefont {H.}~\bibnamefont {Zheng}}, \bibinfo
  {author} {\bibfnamefont {H.}~\bibnamefont {Chen}}, \bibinfo {author}
  {\bibfnamefont {C.}~\bibnamefont {Li}}, \bibinfo {author} {\bibfnamefont
  {Y.}~\bibnamefont {Li}}, \ and\ \bibinfo {author} {\bibfnamefont
  {Y.}~\bibnamefont {Zhang}},\ }\href@noop {} {\bibfield  {journal} {\bibinfo
  {journal} {Opt. Express}\ }\textbf {\bibinfo {volume} {22}},\ \bibinfo
  {pages} {7160} (\bibinfo {year} {2014})}\BibitemShut {NoStop}%
\bibitem [{\citenamefont {Zhang}\ \emph {et~al.}(2017)\citenamefont {Zhang},
  \citenamefont {Huang}, \citenamefont {Conti}, \citenamefont {Wang},
  \citenamefont {Hu}, \citenamefont {Lei},\ and\ \citenamefont
  {Fan}}]{zhangL17a}%
  \BibitemOpen
  \bibfield  {author} {\bibinfo {author} {\bibfnamefont {L.}~\bibnamefont
  {Zhang}}, \bibinfo {author} {\bibfnamefont {P.}~\bibnamefont {Huang}},
  \bibinfo {author} {\bibfnamefont {C.}~\bibnamefont {Conti}}, \bibinfo
  {author} {\bibfnamefont {Z.}~\bibnamefont {Wang}}, \bibinfo {author}
  {\bibfnamefont {Y.}~\bibnamefont {Hu}}, \bibinfo {author} {\bibfnamefont
  {D.}~\bibnamefont {Lei}}, \ and\ \bibinfo {author} {\bibfnamefont {Y.~L.~D.}\
  \bibnamefont {Fan}},\ }\href@noop {} {\bibfield  {journal} {\bibinfo
  {journal} {Opt. Express}\ }\textbf {\bibinfo {volume} {25}},\ \bibinfo
  {pages} {1856} (\bibinfo {year} {2017})}\BibitemShut {NoStop}%
\bibitem [{\citenamefont {Zhang}\ \emph {et~al.}(2018)\citenamefont {Zhang},
  \citenamefont {Pierangeli}, \citenamefont {Conti}, \citenamefont {Fan},\ and\
  \citenamefont {Zhang}}]{zhangX18a}%
  \BibitemOpen
  \bibfield  {author} {\bibinfo {author} {\bibfnamefont {X.}~\bibnamefont
  {Zhang}}, \bibinfo {author} {\bibfnamefont {D.}~\bibnamefont {Pierangeli}},
  \bibinfo {author} {\bibfnamefont {C.}~\bibnamefont {Conti}}, \bibinfo
  {author} {\bibfnamefont {D.}~\bibnamefont {Fan}}, \ and\ \bibinfo {author}
  {\bibfnamefont {L.}~\bibnamefont {Zhang}},\ }\href@noop {} {\bibfield
  {journal} {\bibinfo  {journal} {Opt. Express}\ }\textbf {\bibinfo {volume}
  {26}},\ \bibinfo {pages} {32971} (\bibinfo {year} {2018})}\BibitemShut
  {NoStop}%
\bibitem [{\citenamefont {Bouchet}\ \emph {et~al.}(2018)\citenamefont
  {Bouchet}, \citenamefont {Marsal}, \citenamefont {Sciamanna},\ and\
  \citenamefont {Wolfersberger}}]{bouchet18a}%
  \BibitemOpen
  \bibfield  {author} {\bibinfo {author} {\bibfnamefont {T.}~\bibnamefont
  {Bouchet}}, \bibinfo {author} {\bibfnamefont {N.}~\bibnamefont {Marsal}},
  \bibinfo {author} {\bibfnamefont {M.}~\bibnamefont {Sciamanna}}, \ and\
  \bibinfo {author} {\bibfnamefont {D.}~\bibnamefont {Wolfersberger}},\
  }\href@noop {} {\bibfield  {journal} {\bibinfo  {journal} {Phys. Rev. A}\
  }\textbf {\bibinfo {volume} {97}},\ \bibinfo {pages} {051801} (\bibinfo
  {year} {2018})}\BibitemShut {NoStop}%
\bibitem [{\citenamefont {Thom}(1972)}]{thom_book72a}%
  \BibitemOpen
  \bibfield  {author} {\bibinfo {author} {\bibfnamefont {R.}~\bibnamefont
  {Thom}},\ }\href@noop {} {\emph {\bibinfo {title} {Stabilit\'{e} structurelle
  et morphog\'{e}n\`{e}se : essai d'une th\'{e}orie g\'{e}n\'{e}rale des
  mod\`{e}les.}}}\ (\bibinfo  {publisher} {Reading : W.A. Benjamin},\ \bibinfo
  {year} {1972})\BibitemShut {NoStop}%
\bibitem [{\citenamefont {Berry}\ and\ \citenamefont
  {Upstill}(1980)}]{berry80a}%
  \BibitemOpen
  \bibfield  {author} {\bibinfo {author} {\bibfnamefont {M.~V.}\ \bibnamefont
  {Berry}}\ and\ \bibinfo {author} {\bibfnamefont {C.}~\bibnamefont
  {Upstill}},\ }\href@noop {} {\bibfield  {journal} {\bibinfo  {journal}
  {Progress in Optics}\ }\textbf {\bibinfo {volume} {18}},\ \bibinfo {pages}
  {257} (\bibinfo {year} {1980})}\BibitemShut {NoStop}%
\bibitem [{\citenamefont {Ring}\ \emph {et~al.}(2012)\citenamefont {Ring},
  \citenamefont {Lindberg}, \citenamefont {Mourka}, \citenamefont {Mazilu},
  \citenamefont {Dholakia},\ and\ \citenamefont {Dennis}}]{ring12a}%
  \BibitemOpen
  \bibfield  {author} {\bibinfo {author} {\bibfnamefont {J.~D.}\ \bibnamefont
  {Ring}}, \bibinfo {author} {\bibfnamefont {J.}~\bibnamefont {Lindberg}},
  \bibinfo {author} {\bibfnamefont {A.}~\bibnamefont {Mourka}}, \bibinfo
  {author} {\bibfnamefont {M.}~\bibnamefont {Mazilu}}, \bibinfo {author}
  {\bibfnamefont {K.}~\bibnamefont {Dholakia}}, \ and\ \bibinfo {author}
  {\bibfnamefont {M.~R.}\ \bibnamefont {Dennis}},\ }\href@noop {} {\bibfield
  {journal} {\bibinfo  {journal} {Opt. Express}\ }\textbf {\bibinfo {volume}
  {20}},\ \bibinfo {pages} {18955} (\bibinfo {year} {2012})}\BibitemShut
  {NoStop}%
\bibitem [{\citenamefont {Zang}\ \emph {et~al.}(2019)\citenamefont {Zang},
  \citenamefont {Wang},\ and\ \citenamefont {Li}}]{zangf19a}%
  \BibitemOpen
  \bibfield  {author} {\bibinfo {author} {\bibfnamefont {F.}~\bibnamefont
  {Zang}}, \bibinfo {author} {\bibfnamefont {Y.}~\bibnamefont {Wang}}, \ and\
  \bibinfo {author} {\bibfnamefont {L.}~\bibnamefont {Li}},\ }\href@noop {}
  {\bibfield  {journal} {\bibinfo  {journal} {Results in Physics}\ }\textbf
  {\bibinfo {volume} {15}},\ \bibinfo {pages} {102656} (\bibinfo {year}
  {2019})}\BibitemShut {NoStop}%
\bibitem [{\citenamefont {Zannotti}\ \emph {et~al.}(2017)\citenamefont
  {Zannotti}, \citenamefont {Diebel}, \citenamefont {Boguslawski},\ and\
  \citenamefont {Denz}}]{zannotti17a}%
  \BibitemOpen
  \bibfield  {author} {\bibinfo {author} {\bibfnamefont {A.}~\bibnamefont
  {Zannotti}}, \bibinfo {author} {\bibfnamefont {F.}~\bibnamefont {Diebel}},
  \bibinfo {author} {\bibfnamefont {M.}~\bibnamefont {Boguslawski}}, \ and\
  \bibinfo {author} {\bibfnamefont {C.}~\bibnamefont {Denz}},\ }\href@noop {}
  {\bibfield  {journal} {\bibinfo  {journal} {New J. Phys.}\ }\textbf {\bibinfo
  {volume} {19}},\ \bibinfo {pages} {053004} (\bibinfo {year}
  {2017})}\BibitemShut {NoStop}%
\bibitem [{\citenamefont {Debnath}\ and\ \citenamefont
  {Shah}(2015)}]{debnath_book15a}%
  \BibitemOpen
  \bibfield  {author} {\bibinfo {author} {\bibfnamefont {L.}~\bibnamefont
  {Debnath}}\ and\ \bibinfo {author} {\bibfnamefont {F.~A.}\ \bibnamefont
  {Shah}},\ }\href@noop {} {\emph {\bibinfo {title} {Wavelet Transforms and
  Their Applications}}},\ \bibinfo {edition} {2nd}\ ed.\ (\bibinfo  {publisher}
  {Birkh{\u a}user},\ \bibinfo {year} {2015})\BibitemShut {NoStop}%
\bibitem [{\citenamefont {Baker}\ \emph {et~al.}(2012)\citenamefont {Baker},
  \citenamefont {Jordan},\ and\ \citenamefont {Norris}}]{baker12a}%
  \BibitemOpen
  \bibfield  {author} {\bibinfo {author} {\bibfnamefont {C.~H.}\ \bibnamefont
  {Baker}}, \bibinfo {author} {\bibfnamefont {D.~A.}\ \bibnamefont {Jordan}}, \
  and\ \bibinfo {author} {\bibfnamefont {P.~M.}\ \bibnamefont {Norris}},\
  }\href@noop {} {\bibfield  {journal} {\bibinfo  {journal} {Phys. Rev. B}\
  }\textbf {\bibinfo {volume} {86}},\ \bibinfo {pages} {104306} (\bibinfo
  {year} {2012})}\BibitemShut {NoStop}%
\bibitem [{\citenamefont {Colas}\ and\ \citenamefont
  {Laussy}(2016)}]{colas16a}%
  \BibitemOpen
  \bibfield  {author} {\bibinfo {author} {\bibfnamefont {D.}~\bibnamefont
  {Colas}}\ and\ \bibinfo {author} {\bibfnamefont {F.~P.}\ \bibnamefont
  {Laussy}},\ }\href@noop {} {\bibfield  {journal} {\bibinfo  {journal} {Phys.
  Rev. Lett.}\ }\textbf {\bibinfo {volume} {116}},\ \bibinfo {pages} {026401}
  (\bibinfo {year} {2016})}\BibitemShut {NoStop}%
\bibitem [{\citenamefont {Colas}\ \emph {et~al.}(2018)\citenamefont {Colas},
  \citenamefont {Laussy},\ and\ \citenamefont {Davis}}]{colas18a}%
  \BibitemOpen
  \bibfield  {author} {\bibinfo {author} {\bibfnamefont {D.}~\bibnamefont
  {Colas}}, \bibinfo {author} {\bibfnamefont {F.}~\bibnamefont {Laussy}}, \
  and\ \bibinfo {author} {\bibfnamefont {M.~J.}\ \bibnamefont {Davis}},\
  }\href@noop {} {\bibfield  {journal} {\bibinfo  {journal} {Phys. Rev. Lett.}\
  }\textbf {\bibinfo {volume} {121}},\ \bibinfo {pages} {055302} (\bibinfo
  {year} {2018})}\BibitemShut {NoStop}%
\bibitem [{\citenamefont {Colas}\ \emph {et~al.}(2019)\citenamefont {Colas},
  \citenamefont {Laussy},\ and\ \citenamefont {Davis}}]{colas19a}%
  \BibitemOpen
  \bibfield  {author} {\bibinfo {author} {\bibfnamefont {D.}~\bibnamefont
  {Colas}}, \bibinfo {author} {\bibfnamefont {F.}~\bibnamefont {Laussy}}, \
  and\ \bibinfo {author} {\bibfnamefont {M.~J.}\ \bibnamefont {Davis}},\
  }\href@noop {} {\bibfield  {journal} {\bibinfo  {journal} {Phys. Rev. B}\
  }\textbf {\bibinfo {volume} {99}},\ \bibinfo {pages} {214301} (\bibinfo
  {year} {2019})}\BibitemShut {NoStop}%
\bibitem [{not({\natexlab{a}})}]{note5}%
  \BibitemOpen
  \href@noop {} {} ({\natexlab{a}}),\ \bibinfo {note} {one can compute the
  total normalisation from Eq.~(2) as $N=\int_{-\infty}^{+\infty} |\psi_0(x)|^2
  dx =\exp(\frac{2a^3}{3b^3})/8\sqrt{2\pi a b}$ and this result is
  finite.}\BibitemShut {Stop}%
\bibitem [{\citenamefont {Wertz}\ \emph {et~al.}(2010)\citenamefont {Wertz},
  \citenamefont {Ferrier}, \citenamefont {Solnyshkov}, \citenamefont {Johne},
  \citenamefont {Sanvitto}, \citenamefont {Lema\^itre}, \citenamefont {Sagnes},
  \citenamefont {Grousson}, \citenamefont {Kavokin}, \citenamefont
  {Senellart},\ and\ \citenamefont {an~J.~Bloch}}]{wertz10a}%
  \BibitemOpen
  \bibfield  {author} {\bibinfo {author} {\bibfnamefont {E.}~\bibnamefont
  {Wertz}}, \bibinfo {author} {\bibfnamefont {L.}~\bibnamefont {Ferrier}},
  \bibinfo {author} {\bibfnamefont {D.~D.}\ \bibnamefont {Solnyshkov}},
  \bibinfo {author} {\bibfnamefont {R.}~\bibnamefont {Johne}}, \bibinfo
  {author} {\bibfnamefont {D.}~\bibnamefont {Sanvitto}}, \bibinfo {author}
  {\bibfnamefont {A.}~\bibnamefont {Lema\^itre}}, \bibinfo {author}
  {\bibfnamefont {I.}~\bibnamefont {Sagnes}}, \bibinfo {author} {\bibfnamefont
  {R.}~\bibnamefont {Grousson}}, \bibinfo {author} {\bibfnamefont {A.~V.}\
  \bibnamefont {Kavokin}}, \bibinfo {author} {\bibfnamefont {P.}~\bibnamefont
  {Senellart}}, \ and\ \bibinfo {author} {\bibfnamefont {G.~M.}\ \bibnamefont
  {an~J.~Bloch}},\ }\href@noop {} {\bibfield  {journal} {\bibinfo  {journal}
  {Nat. Phys.}\ }\textbf {\bibinfo {volume} {6}},\ \bibinfo {pages} {860}
  (\bibinfo {year} {2010})}\BibitemShut {NoStop}%
\bibitem [{\citenamefont {Sonin}(2016)}]{sonin_book16a}%
  \BibitemOpen
  \bibfield  {author} {\bibinfo {author} {\bibfnamefont {E.~B.}\ \bibnamefont
  {Sonin}},\ }\href@noop {} {\emph {\bibinfo {title} {Dynamics of Quantised
  Vortices in Superfluids}}}\ (\bibinfo  {publisher} {Cambridge University
  Press},\ \bibinfo {year} {2016})\BibitemShut {NoStop}%
\bibitem [{\citenamefont {Kittel}(1996)}]{kittel_book96a}%
  \BibitemOpen
  \bibfield  {author} {\bibinfo {author} {\bibfnamefont {C.}~\bibnamefont
  {Kittel}},\ }\href@noop {} {\emph {\bibinfo {title} {Introduction to Solid
  State Physics}}},\ \bibinfo {edition} {11th}\ ed.\ (\bibinfo  {publisher}
  {Wiley},\ \bibinfo {year} {1996})\BibitemShut {NoStop}%
\bibitem [{not({\natexlab{b}})}]{note7}%
  \BibitemOpen
  \href@noop {} {} ({\natexlab{b}}),\ \bibinfo {note} {see Supplemental
  Material at [URL will be inserted by publisher] for a wavelet-based analysis
  of the diffusion of a Gaussian wave packet, further discussion of the
  Airy-engineered wave front, and movies providing time-animated versions of
  Figures 1, 2 and 3.}\BibitemShut {Stop}%
\bibitem [{not({\natexlab{c}})}]{note2}%
  \BibitemOpen
  \href@noop {} {} ({\natexlab{c}}),\ \bibinfo {note} {we note that the WT is
  governed by an uncertainty principle between its resolution in space
  ($\Delta_x$) and momentum ($\Delta_k$), so that the product $\Delta_x
  \Delta_k$ cannot be arbitrarily small. One typically needs to adapt the WT
  parameters such as the wavelet frequency to obtain a suitable resolution in
  the desired momentum/space range. In Fig.~\ref{fig:1}(c--f) the WT is less
  adapted for the low momenta, which leads to broader energy distributions in
  this momentum range.}\BibitemShut {Stop}%
\bibitem [{\citenamefont {Hu}\ \emph {et~al.}(2012)\citenamefont {Hu},
  \citenamefont {Sun}, \citenamefont {Bongiovanni}, \citenamefont {Song},
  \citenamefont {Lou}, \citenamefont {Xu}, \citenamefont {Chen},\ and\
  \citenamefont {Morandotti}}]{HuY12a}%
  \BibitemOpen
  \bibfield  {author} {\bibinfo {author} {\bibfnamefont {Y.}~\bibnamefont
  {Hu}}, \bibinfo {author} {\bibfnamefont {Z.}~\bibnamefont {Sun}}, \bibinfo
  {author} {\bibfnamefont {D.}~\bibnamefont {Bongiovanni}}, \bibinfo {author}
  {\bibfnamefont {D.}~\bibnamefont {Song}}, \bibinfo {author} {\bibfnamefont
  {C.}~\bibnamefont {Lou}}, \bibinfo {author} {\bibfnamefont {J.}~\bibnamefont
  {Xu}}, \bibinfo {author} {\bibfnamefont {Z.}~\bibnamefont {Chen}}, \ and\
  \bibinfo {author} {\bibfnamefont {R.}~\bibnamefont {Morandotti}},\
  }\href@noop {} {\bibfield  {journal} {\bibinfo  {journal} {Opt. Lett.}\
  }\textbf {\bibinfo {volume} {37}},\ \bibinfo {pages} {3201} (\bibinfo {year}
  {2012})}\BibitemShut {NoStop}%
\bibitem [{\citenamefont {Egorov}\ \emph {et~al.}(2009)\citenamefont {Egorov},
  \citenamefont {Skryabin}, \citenamefont {Yulin},\ and\ \citenamefont
  {Lederer}}]{egorov09a}%
  \BibitemOpen
  \bibfield  {author} {\bibinfo {author} {\bibfnamefont {O.~A.}\ \bibnamefont
  {Egorov}}, \bibinfo {author} {\bibfnamefont {D.~V.}\ \bibnamefont
  {Skryabin}}, \bibinfo {author} {\bibfnamefont {A.~V.}\ \bibnamefont {Yulin}},
  \ and\ \bibinfo {author} {\bibfnamefont {F.}~\bibnamefont {Lederer}},\
  }\href@noop {} {\bibfield  {journal} {\bibinfo  {journal} {Phys. Rev. Lett.}\
  }\textbf {\bibinfo {volume} {102}},\ \bibinfo {pages} {153904} (\bibinfo
  {year} {2009})}\BibitemShut {NoStop}%
\bibitem [{\citenamefont {Sich}\ \emph {et~al.}(2011)\citenamefont {Sich},
  \citenamefont {Krizhanovskii}, \citenamefont {Skolnick}, \citenamefont
  {Gorbach}, \citenamefont {Hartley}, \citenamefont {Skryabin}, \citenamefont
  {Cerda-M\'endez}, \citenamefont {Biermann}, \citenamefont {Hey},\ and\
  \citenamefont {Santos}}]{sich11a}%
  \BibitemOpen
  \bibfield  {author} {\bibinfo {author} {\bibfnamefont {M.}~\bibnamefont
  {Sich}}, \bibinfo {author} {\bibfnamefont {D.}~\bibnamefont {Krizhanovskii}},
  \bibinfo {author} {\bibfnamefont {M.}~\bibnamefont {Skolnick}}, \bibinfo
  {author} {\bibfnamefont {A.}~\bibnamefont {Gorbach}}, \bibinfo {author}
  {\bibfnamefont {R.}~\bibnamefont {Hartley}}, \bibinfo {author} {\bibfnamefont
  {D.~V.}\ \bibnamefont {Skryabin}}, \bibinfo {author} {\bibfnamefont {E.~A.}\
  \bibnamefont {Cerda-M\'endez}}, \bibinfo {author} {\bibfnamefont
  {K.}~\bibnamefont {Biermann}}, \bibinfo {author} {\bibfnamefont
  {R.}~\bibnamefont {Hey}}, \ and\ \bibinfo {author} {\bibfnamefont
  {P.}~\bibnamefont {Santos}},\ }\href@noop {} {\bibfield  {journal} {\bibinfo
  {journal} {Nat. Photon.}\ }\textbf {\bibinfo {volume} {6}},\ \bibinfo {pages}
  {50} (\bibinfo {year} {2011})}\BibitemShut {NoStop}%
\bibitem [{not({\natexlab{d}})}]{note8}%
  \BibitemOpen
  \href@noop {} {} ({\natexlab{d}}),\ \bibinfo {note} {the function is named
  after Trevor Pearcey who, at the end of World War II, first obtained the
  numerical values for $\textrm{Pe}(x,y)$ in a computational tour de force
  using a mechanical differential analyser at the University of
  Cambridge.}\BibitemShut {Stop}%
\bibitem [{not({\natexlab{e}})}]{note9}%
  \BibitemOpen
  \href@noop {} {} ({\natexlab{e}}),\ \bibinfo {note} {a closed form for
  $\mathscr{F}_{k} [\psi_0(x)]$, \textit{i.e.} including the truncation, cannot
  be obtained in terms of simple functions. We thus based our analysis on the
  decomposition the infinite energy version of the beam ($\sigma =0$) which we
  expect to be a good approximation for the FEPB, since the cut-off is chosen
  as $\sigma \ll 1$.}\BibitemShut {Stop}%
\end{thebibliography}
\end{document}